# Softwarization of Internet of Things Infrastructure for Secure and Smart Healthcare

Mohammad A. Salahuddin, *Member, IEEE*, Ala Al-Fuqaha, *Senior Member, IEEE*, Mohsen Guizani, *Fellow, IEEE*, Khaled Shuaib, *SeniorMember, IEEE*, and Farag Sallabi, *Member, IEEE*

*Abstract*—We propose an agile softwarized infrastructure for flexible, cost effective, secure and privacy preserving deployment of Internet of Things (IoT) for smart healthcare applications and services. It integrates state-of-the-art networking and virtualization techniques across IoT, fog and cloud domains, employing Blockchain, Tor and message brokers to provide security and privacy for patients and healthcare providers. We propose a novel platform using Machine-to-Machine (M2M) messaging and rule-based beacons for seamless data management and discuss the role of data and decision fusion in the cloud and the fog, respectively, for smart healthcare applications and services.

## I. INTRODUCTION

SMART HEALTHCARE will be the most dominant Internet of Things (IoT) application [1], optimizing healthcare delivery and experience while minimizing operational and capital expenditure (OPEX/CAPEX) by leveraging cloud and fog computing for healthcare providers. Smart healthcare applications and services require collection, aggregation and analysis of raw sensory data [1]. The challenge lies in aggregating heterogeneous data from different types of sensors and other sources. The data characteristics of IoT for healthcare are accurately captured by Big Data characteristics of volume, velocity and variety. The numerous, ambient and embedded devices in our environment generate voluminous data ranging in orders of magnitudes (e.g. kilobytes to gigabytes). The voluminous data and Smart City applications and services require processing times that range from batch, to pseudo-real, to real-time processing of various types of data (e.g. text, audio, video, etc.) To this end, we propose an agile softwarized infrastructure that embraces cloud and fog computing, Blockchain, Tor and message brokers for flexible, cost effective, secure and privacy preserving deployment of IoT for smart healthcare applications and services.

Specifically, our contributions can be delineated as (i) a system architecture of a state-of-the-art softwarized IoT infrastructure for smart healthcare, (ii) a novel platform with Machine-to-Machine (M2M) messaging and rule-based beacons for seamless data management for smart healthcare applications and services, and (iii) analyzing the role of data and decision fusion to facilitate smart healthcare applications and services.

## II. System Architecture

In Fig. 1, we illustrate the interplay of the various IoT elements and the networking and computing technologies that implement the softwarized infrastructure for smart healthcare. The IoT elements include smart sensors, of different size and types, that can monitor biological, chemical and physical parameters, and process and record raw sensory data. Transceivers on the sensors enable communication with base stations through wireless interface. The more powerful base stations act as data aggregators, sink nodes, gateways to the Internet and cloud or IoT gateways [2] for smart healthcare applications and services. The IoT gateways overcome the heterogeneity in devices and network protocols and enable connectivity [3].

### A. Softwarization

Generally, sensor networks are application-oriented silos that lack dynamic configurability. Software-Defined Networking (SDN) offers a cost effective solution for improving agility and flexibility of sensor networks. SDN decouples the control and data planes in a network and allows dynamic and flexible configuration and management of data forwarding rules. This increases interoperability between communication protocols [4] and reduces cost of network deployment, configuration and management by easily updating commercial-off-the-shelf (COTS) hardware to become an SDN-compliant element [5].The softwarized infrastructure also benefits from programmability of COTS hardware to perform network functions and even delivery of end-to-end services[6], via Network Functions Virtualization (NFV).In the softwarized infrastructure, the virtual network functions (VNFs) are chained together to compose a service, while the SDN controller facilitates traffic steering between virtual and physical network functions and applications. A software NFV manager and orchestrator (MANO) is used for creating, configuring, managing and monitoring the VNFs in the Network Functions Virtualization Infrastructure (NFVI). The IoT gateways are directly or indirectly connected to the virtualized nodes that promote agility and cost effective delivery of applications and services.

The cloud, in Fig. 1, offers the opportunity to optimize healthcare delivery [7] with analytics, reduce OPEX/CAPEX and increase security and privacy of data, applications and services hosted in the cloud (explained later). The cloud maintains comprehensive patient records and augments it with patients' biometric [8], genomic [8], familial [8] and social data [9], giving healthcare providers a holistic perspective of the mental, physical and social health of the patients. Data analytics enable early detection and prevention of projected risks for patients, identify possible epidemic or outbreaks in population, improve precision in healthcare delivery and identify wastage and misuse of healthcare resources to reduce OPEX/CAPEX.

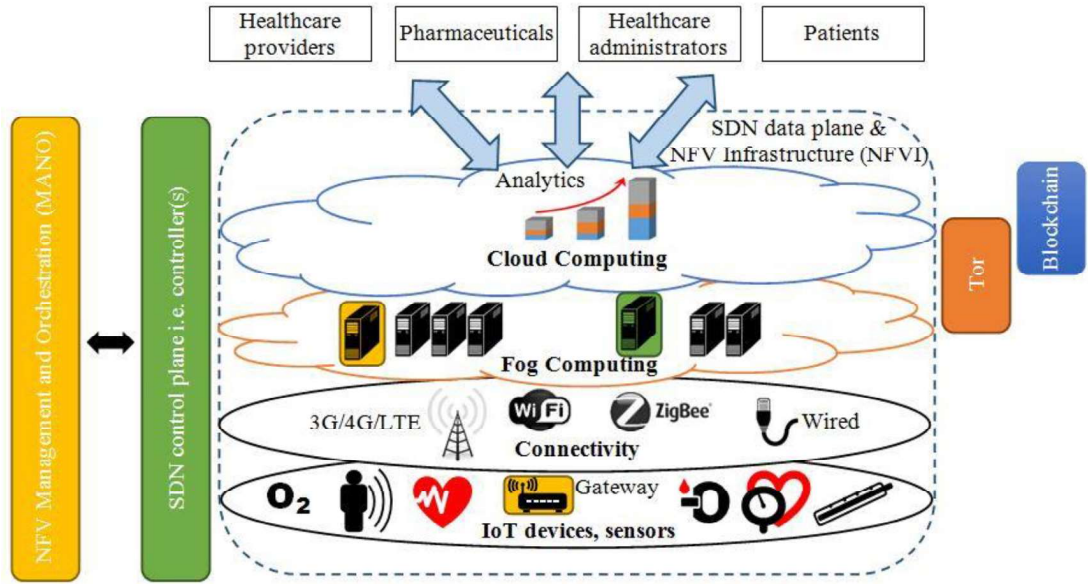

Fig. 1. System architecture of IoT for smart healthcare.

## B. Security and Privacy

The security and privacy measures offered by the cloud increase the resiliency of data. The privacy of patient records and other confidential and sensitive healthcare data is ensured using mechanisms like the Onion Routing (Tor) ([10], [11]) in tandem with Machine-to-Machine (M2M) message protocols, such as the Message Queue Telemetry Transport protocol (MQTT) to preserve anonymity of online users and data. Essentially, it enables security against network surveillance threats to gain access to sensitive patient and hospital records.

Tor removes mapping between user IP address, identification and services or servers being used to achieve user anonymity to prevent possible tracking. This is accomplished by imposing an overlay network of secure connections between nodes in the underlying network through which random paths are selected for communication. Since, Tor can introduce delay [12] and unpredictability, we propose to employ it between fog nodes and the cloud. Fog nodes are more localized and can meet the stringent real-time requirements, in contrast to Tor in the cloud. Therefore, it is important to strike a tradeoff between anonymity and latency when using Tor depending on the application.

After preserving data of patients and hospitals, Blockchain can guarantee security of sensitive data by tracking and authorizing access to confidential medical records, as illustrated in Fig. 2. Blockchain serve as a distributed database that hardens medical reports against tampering [13]. As a distributed trust mechanism, Blockchain address security issues associated with a deployed distributed database of patient records. These could be managed by different parties, such as caregivers, hospitals, pharmacies, insurance companies, regulators and the patients. Blockchain relies on public key cryptography to track trusted transactions pertaining to distributed patient records while validating and preserving confidentiality, integrity and availability. This ensures that records are not lost or being wrongly modified, falsified or accessed by unauthorized users. Smart contracts easily, transparently and seamlessly transfer patient information among the members of the Blockchain via self-executing agreed upon conditions trusted by all members.

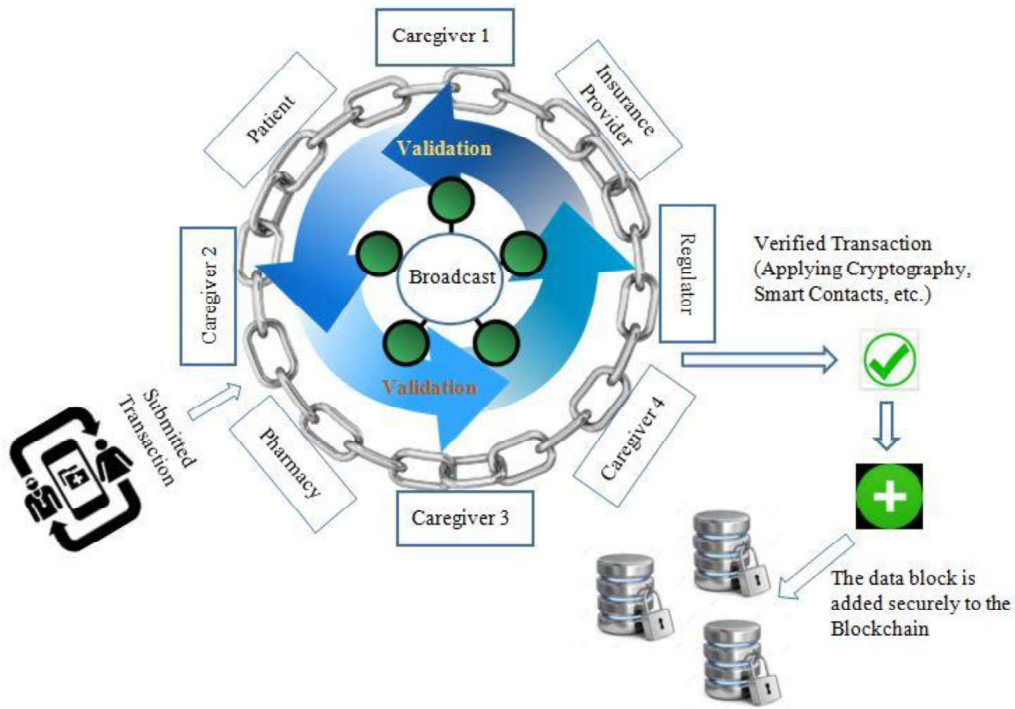

Fig. 2. Security via Blockchain in softwarized IoT for smart healthcare.

Healthcare applications and services cannot afford the latency of the cloud, therefore, fog computing brings cloud-like resources and computing closer to the end-users. The fog nodes in Fig. 1, are smaller than cloud nodes, but more powerful than the IoT devices and gateways. It offers low latency and high performance to process and aggregate localized data and reduces unnecessary traffic to the cloud. Data collected from the sensor network in the IoT for healthcare must be aggregated for analytics and healthcare applications and services.

III. DATA AGGREGATION

In this section, we illustrate the role of data and decision in data management with a specific use case for IoT in smart healthcare. Consider, two small sensor networks composed of an IoT for cardio health monitoring of a patient in a healthcare facility. The cardio health monitoring application requires real-time monitoring and alerts and logging of patient data in the cloud for detailed analytics for quality and cost of healthcare. One sensor network consists of electro cardio graph (ECG) sensors, while the other consists of photo plethysmography (PPG) sensors. Each sensor network has a gateway. The gateway in the ECG sensor network is also an IoT gateway and a fog node connected to a healthcare database and server in the cloud.

The cardio health monitoring application requires collaboration amongst the ECG and PPG sensors for fusion of data and, or information to accurately and efficiently monitor and control, by remote medicine administration, the cardio health of the patient. There are two fundamental approaches to data processing, data or decision fusion.

## A. Data Processing

In decision fusion, smart sensors locally process the individual sensor(s) measurements and compute the decision for a criterion, for instance, abnormal heart beat rate. In data fusion, the sensors transmit raw sensory data, such as, voltage, light absorption, etc., to a base station or predetermined gateway for decision processing, for example, normal heat beat rate. In decision fusion, the raw sensory data is processed locally at the sensor and the information (decision) about the raw sensory data, such as, heart rate based on electric voltage measured by ECG sensor, is transmitted to the ECG sensor gateway. The PPG sensor gateway collects the flow rate and transmits the information to the IoT gateway, in our case the ECG sensor gateway. At the IoT gateway, in the fog, the decision regarding cardiac health of the patient is aggregated, processed and logged in healthcare servers in the cloud for data analytics and reported to authorized healthcare providers and personnel on their mobile devices. In data fusion, the raw sensory data is transmitted to sensor network gateways, where it is filtered and processed into information and transmitted to IoT gateway for further logging and reporting.

There are various advantages and disadvantages of data and decision fusion in IoT for healthcare. Data fusion consumes more bandwidth for transmitting raw sensory data, which could have many features with high dimensionality, e.g. data from image sensors used in endoscopy. Whereas, decision fusion requires less bandwidth since decision, that is refined information, is transmitted over the network.

Decision fusion results in loss of accuracy, since raw sensory data processing will contain inherent rounding errors in computation based on processing accuracy of sensors. On the other hand, data fusion yields more accurate results since the computations are performed on the more powerful sensor network gateways. Consequentially, decision fusion consumes the constrained power of the sensors to compute and transmit the information, whereas, in data fusion sensors consume more power to transmit high dimensional data. Since, radio transmission consumes more power than processor computations.

## B. Agile platform for IoT

Evidently, there are various levels of decision and data fusion within an IoT-enabled smart healthcare. We propose a solution that utilizes a deep field-programmable gate array (FPGA) that consists of hardware and software components, which deliver location and context aware services using M2M communications, as illustrated in Fig. 3. Our platform, called flexBeacon, offers a unified hardware, so that almost any sensor or actuator can be connected to the board facilitating easy sharing of telemetry and access to remote control services. The hardware abstraction of flexBeacon allows healthcare providers to configure rules and data flow models to design healthcare monitoring and controlling applications and services to customize healthcare experience.

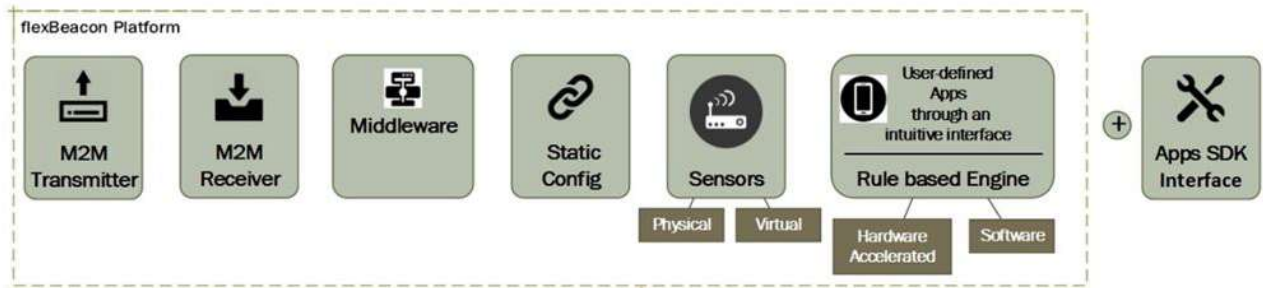

Fig. 3. IoT rule-based beacon platform.

flexBeacon allows users to define rules and logic to control the flow of data to design monitoring and actuation controlling applications. It enables seamless data aggregation and analytics to streamline decision making for patients, healthcare providers and administrators of healthcare facilities. The M2M based communication and FPGA based hardware enables low-latency and high performance of user defined rules and flow models for data collection, aggregation, correlation and reporting.

The obvious advantages of our proposed platform include, seamless data aggregation, management without loss of accuracy and consumption of constrained resources, while greatly reducing the cost of a softwarized IoT for smart healthcare.

## IV. Conclusion and Challenges

Smart healthcare applications and services can perform real-time monitoring and actuation for healthcare needs of patients and use data analytics in the cloud to improve the quality of healthcare and experience for patients while reducing cost of providing healthcare services.

To this end, we delineate the system architecture of a state-of-the-art IoT infrastructure that is agile, flexible, secure and privacy preserving for smart healthcare, which can be deployed cost effectively. We also envisage a novel FPGA platform for high performance, low latency and local execution of end-user defined rules for beacon and flow, M2M transceiver and microcontroller for seamless integration of data for agile deployment of smart healthcare applications and services. Furthermore, we discuss the role of data and decision fusion for data management in IoT-based smart healthcare.

However, there are various challenges in realizing this secure and agile IoT infrastructure for smart healthcare. Some of these challenges include:

1. Softwarization of IoT
    i. There is a need for seamless integration of 5G systems with a softwarized IoT. Researchers have evaluated the performance of existing softwarization technologies, such as SDN and NFV with 5G systems [14]. Obvious implications of SDN and NFV for 5G includes resource, spectrum and transmission power management, optimal connection between network devices, transceivers and physical elements and providing services with different QoS.
    ii. To facilitate a softwarized IoT, various aspects of SDN and NFV have to be further scrutinized. Including, but not limited to, defining the key performance indicators (KPIs) for gauging the performance of softwarized network elements, functions and applications, designing and managing distributed controllers in the control plane and distributed network functions to ensure vertical and horizontal scalability, autonomous orchestration of network functions and services across the softwarized middleware.

2. Security and Privacy
    i. The protection against traffic analysis to prevent data forensics by inference and to improve obfuscation while maintaining accountability and the privacy of individual transactions. This needs further attention as all transactions can be seen by every member on the Blockchain. To this end, secure communication protocols between individual IoT devices or members on the Blockchain can be used. In addition, the use of homomorphic encryption techniques and zero knowledge proofs [15] can also be utilized as seen applicable depending on the availability of resources and the technical capabilities of the participating IoT devices.
    ii. Proper and logical implementations of Blockchains based on smart contracts enforced legally is an essential aspect of a successful deployment at a large scale. This will improve system performance and minimize blocked transactions due to consensus not being reached. The integration of legal contracts as part of the deployed smart contracts can be used for proper enforcement and to control any misbehaving members. This will require the use of hashing techniques where a generated hash of the legal contract can be made part of the smart contract. This will ensure the confidentiality of used legal contracts while being able to resolve any disputes.

ACKNOWLEDGEMENT

This article was made possible by NPRP grant #[7-1113-1-199] from the Qatar National Research Fund (a member of Qatar Foundation). The statements made herein are solely the responsibility of the authors.

*Mohammad A. Salahuddin* is with the David R. Cheriton School of Computer Science, University of Waterloo, Waterloo, ON, N2L 3G1, Canada. **(e-mail: mohammad.salahuddin@ieee.org)**

*Ala Al-Fuqaha* is with the Department of Computer Science, Western Michigan University, Kalamazoo, MI, 49008, USA. **(email: ala.alfuqaha@wmich.edu)**

*Mohsen Guizani* is the Chair of the Electrical and Computer Engineering Department, University of Idaho, Moscow, ID, 83844, USA. **(email: mguizani@uidaho.edu)**

*Khaled Shuaib* and *Farag Sallabi* are with the College of Information Technology, United Arab Emirates University, Al Ain, UAE. **(emails: {k.shuaib, f.sallabi}@uaeu.ac.ae)**